\documentclass[useAMS,usenatbib]{mn2e}
\usepackage{graphics,rotating,multirow,bm,color,hyperref}
\hypersetup{
    colorlinks=true,
    linkcolor=black,
    citecolor=black,
}

\title[Extended Excursion Set Approach for Chameleon Models]
  {An Extended Excursion Set Approach to  Structure Formation in Chameleon Models}
\author[Baojiu~Li and George~Efstathiou]
  {Baojiu~Li$^{1,2,3,4,}$\thanks{E-mail: b.li@damtp.cam.ac.uk}, George~Efstathiou$^{2,3,}$\thanks{E-mail: gpe@ast.cam.ac.uk}\\
  $^1$DAMTP, Centre for Mathematical Sciences, University of Cambridge, Wilberforce Road, Cambridge CB3 0WA, UK\\
  $^2$Kavli Institute for Cosmology Cambridge, Madingley Road, Cambridge CB3 0HA, UK\\
  $^3$Institute of Astronomy, University of Cambridge, Madingley Road, Cambridge CB3 0HA, UK\\
  $^4$Institute for Computational Cosmology, Department of Physics, University of Durham, South Road, Durham DH1 3LE, UK}
  
\pagerange{\pageref{firstpage}--\pageref{lastpage}} \pubyear{2011}

\def\LaTeX{L\kern-.36em\raise.3ex\hbox{a}\kern-.15em
    T\kern-.1667em\lower.7ex\hbox{E}\kern-.125emX}

\newcommand{\tcr}[1]{\textcolor{black}{#1}}

\begin{document}

\label{firstpage}

\maketitle

\begin{abstract}
In attempts to explain dark energy, a number of models have been
proposed in which the formation of large-scale structure depends on
the local environment. These models are highly non-linear and
difficult to analyse analytically.  $N$-body simulations have
therefore been used to study their  non-linear evolution.
Here we extend excursion set theory to incorporate environmental
effects on structure formation. We apply the method to a chameleon
model and calculate observables such as the non-linear mass function
at various redshifts.  The method can be generalized to study other
obervables and other models of environmentally dependent
interactions. The analytic methods described here should prove useful
in delineating which models deserve more detailed study with $N$-body
simulations.
\end{abstract}


\section{Itntroduction}

\label{sect:intro}

One of the most challenging questions in contemporary physics is the
nature of the dark energy, which is believed to be driving
the accelerating expansion of the Universe \citep{riess1998,
  perlmutter1999}. \citet{cst2006} present a comprehensive
review of theoretical models to explain the apparent acceleration
of the Universe. However, at present there is no compelling evidence
for any new physics other than the addition of a cosmological
constant to the Einstein field equations. 

Models of dark energy can be broadly placed into two categories.  In
the first, the dark energy affects the expansion rate of the Universe
but does not interact directly with the dark matter. Examples of this
type of model include the standard $\Lambda$CDM paradigm and
quintessence models \citep{wcos2000}. In the second category, the dark
energy and matter (both dark and baryonic) interact with each other
with an interaction strength which may depend on the local
environment. Examples include the chameleon coupled scalar field model
\citep{kw, ms}, $f(R)$ gravity \citep{cddatt2005}, the environmentally
dependent dilaton model \citep{bbds2010} and also the symmetron model
\citep{hk2010}.

The possibility of environmentally dependent interactions needs to be
considered when relating laboratory measurements to cosmological
scales. Consider, for example, a  scalar field coupled to matter.
The scalar field could mediate a `fifth force' between matter
particles. Current laboratory experiments and Solar System tests have
shown that such a fifth force must be either extremely weak or
short-range (less than about a millimetre) \citep{cw2006}. 
However, it is  possible that the strength
and range of the fifth force depend on the environment so that
locally, where the matter density is high, it is strongly suppressed,
and it is restored in empty environments. In this
situation, laboratory experiments cannot constrain
a fifth force that may have observational consequences on
cosmological scales.

Analytical models of structure formation on galaxy and cluster scales
are notoriously difficult even in the case of standard Newtonian
gravity. The evolution of structure in models with environmentally
dependent interactions is even more complicated because the fifth
force itself is highly non-linear. Consequently, studies so far have
relied on full $N$-body simulations \citep{lz2009, lz2010, lb2011, li2011, lmb2011, lztk2011, zlk2011, bbdls2011, dlmw2011, oyaizu2008, olh2008, sloh2009}.

However, large $N$-body simulations require supercomputing resources
and  are time consuming. They can be justified for testing physically
well motivated models such as the $\Lambda$CDM model, which contains
few parameters, many of which are now well constrained experimentally, see
e.g., \citet{wmap7}.  Models with a fifth force, on the other hand,
span a wide parameter space reflecting our lack of knowledge of the
underlying physics. It is difficult to sample a large parameter space
using full $N$-body simulations, hence the need for an analytic
description of structure formation that can, at least, isolate regions
of parameter space that merit further investigation using simulations.

Semi-analytical models, such as excursion set theory (see
\cite{zentner2007} for a recent review), have been developed as
alternatives to full $N$-body simulations and shown to agree with the
latter well. The excursion set approach has been generalised to some
non-standard structure-formation scenarios \citep{mss2009, phs2011}.
However, these studies do not consider the case of environmentally
dependent interactions.

The aim of this paper is to generalise the excursion set approach to
take account of environmentally dependent interactions explicitly. As
we will see, non-linear collapse of structures could be very different
in different environments, and indeed the environments themselves
evolve in time as well. We will first specify the environments \tcr{using what we call the fixed-scale approximation}, then
use a simplified model to study spherical collapse within these
environments. We then calculate observable properties by averaging
over the distribution of environments. In this paper we have chosen
the chameleon model as a working example, but the methods developed
are more general and with suitable changes can be applied to other
models with environmentally dependent interactions. The theoretical
framework developed here can therefore be used to quickly estimate the
parameter ranges of any specific theory that may have interesting
(and potentially testable) consequences on structure formation.

The layout of this paper is as follows. We introduce the basic
formulae for a chameleon-like coupled scalar field (our working
example) in \S~\ref{sect:model} and summarise the spherically
symmetric solutions which will be used later to study the spherical
collapse of overdensities. \S~\ref{sect:method} presents the main
results of this paper.  We introduce the traditional excursion set
theory in \S~\ref{subsect:excursion} and in \S~\ref{subsect:chameleon}
we show how the environmental dependence in the chameleon model can be
approximated using only two variables.  \S~\ref{subsect:collapse}
describes a generalised spherical collapse model in which an
overdensity collapses inside an evolving environment. Finally in \S~\ref{sect:applications} we make an application of the generalized excursion method to a range of chameleon models. Our conclusions are summarized in
\S~\ref{sect:summary}.

\section{The Theoretical Model}

\label{sect:model}

This section lays down the theoretical framework for 
investigating the effects
of coupled scalar field(s) in cosmology.  We present the
relevant general field equations in \S~\ref{subsect:equations},
specify the models analysed in this paper  in
\S~\ref{subsect:specification}, and then briefly summarise the
spherically symmetric solutions in
\S~\ref{subsect:spherical}.

\subsection{Cosmology with a Coupled Scalar Field}

\label{subsect:equations}

The equations presented in this sub-section  are derived
and discussed in \citep{lz2009, lz2010, lb2011}.
They will be used extensively in the rest of 
this paper and  are presented here for completeness
and to establish the notation used in later sections.

We start from a Lagrangian density
\begin{eqnarray}\label{eq:lagrangian}
{\cal L} =
{1\over{2}}\left[{R\over\kappa}-\nabla^{a}\varphi\nabla_{a}\varphi\right]
+V(\varphi) - C(\varphi){\cal L}_{\rm{DM}} +
{\cal L}_{\rm{S}},
\end{eqnarray}
in which $R$ is the Ricci scalar, $\kappa=8\pi G$ with $G$ being the
gravitational constant, ${\cal L}_{\rm{DM}}$ and
${\cal L}_{\rm{S}}$ are respectively the Lagrangian
densities for dark matter and standard model fields. $\varphi$ is
the scalar field and $V(\varphi)$ its potential; the coupling
function $C(\varphi)$ characterises the coupling between $\varphi$
and dark matter. Given the functional forms for $V(\varphi)$ and $C(\varphi)$
a coupled scalar field model is then fully specified.

Varying the total action with respect to the metric $g_{ab}$, we
obtain the following expression for the total energy momentum
tensor in this model:
\begin{eqnarray}\label{eq:emt}
T_{ab} = \nabla_a\varphi\nabla_b\varphi -
g_{ab}\left[{1\over2}\nabla^{c}\nabla_{c}\varphi-V(\varphi)\right]+ C(\varphi)T^{\rm{DM}}_{ab} + T^{\rm{S}}_{ab},
\end{eqnarray}
where $T^{\rm{DM}}_{ab}$ and $T^{\rm{S}}_{ab}$ are the
energy momentum tensors for (uncoupled) dark matter and standard
model fields. The existence of the scalar field and its coupling
change the form of the energy momentum tensor leading to
potential changes in the background cosmology and 
structure formation.

The coupling to a scalar field produces a direct
interaction (fifth force) between dark matter
particles due to the exchange of scalar quanta. This is best
illustrated by the geodesic equation for dark matter particles
\begin{eqnarray}\label{eq:geodesic}
{{d^{2}\bf{r}}\over{dt^2}} = -\vec{\nabla}\Phi -
{{C_\varphi(\varphi)}\over{C(\varphi)}}\vec{\nabla}\varphi,
\end{eqnarray}
where $\bf{r}$ is the position vector, $t$ the (physical) time, $\Phi$
the Newtonian potential and $\vec{\nabla}$ is the spatial
derivative. $C_\varphi\equiv dC/d\varphi$. The second term in the
right hand side is the fifth force and only exists for coupled matter
species (dark matter in our model). The fifth force also changes the
clustering properties of the dark matter. 

To solve the above two equations we need to know both the time
evolution and the spatial distribution of $\varphi$, {\it i.e.}  we
need the solutions to the scalar field equation of motion (EOM)
\begin{eqnarray}
\nabla^{a}\nabla_a\varphi + {{dV(\varphi)}\over{d\varphi}} +
\rho_{\rm{DM}}{\frac{dC(\varphi)}{d\varphi}} = 0,
\end{eqnarray}
or equivalently
\begin{eqnarray}
\nabla^{a}\nabla_a\varphi + {{dV_{eff}(\varphi)}\over{d\varphi}} =
0,
\end{eqnarray}
where we have defined
\begin{eqnarray}
V_{eff}(\varphi) = V(\varphi) + \rho_{\rm{DM}}C(\varphi).
\end{eqnarray}
The background evolution of $\varphi$ can be solved easily given
the present day value of  $\rho_{\rm{DM}}$ since 
$\rho_{\rm{DM}}\propto a^{-3}$. We can then divide $\varphi$
into two parts, $\varphi=\bar{\varphi}+\delta\varphi$, where
$\bar{\varphi}$ is the background value and $\delta\varphi$ is its
(not necessarily small nor linear) perturbation, and subtract the
background part of the scalar field equation of motion from the full equation
to obtain the equation of motion for $\delta\varphi$. In the
quasi-static limit in which we can neglect time derivatives of
$\delta\varphi$ as compared with its spatial derivatives (which
turns out to be a good approximation on \tcr{galactic and cluster} scales),
we find
\begin{eqnarray}\label{eq:scalar_eom}
\vec{\nabla}^{2}\varphi =
{{dC(\varphi)}\over{d\varphi}}\rho_{\rm{DM}} -
{{dC(\bar{\varphi})}\over{d\bar{\varphi}}}\bar{\rho}_{\rm{DM}} +
{{dV(\varphi)}\over{d\varphi}} -
{{dV(\bar{\varphi})}\over{d\bar{\varphi}}},
\end{eqnarray}
where $\bar{\rho}_{\rm{DM}}$ is the background dark matter
density.

The computation of the scalar field $\varphi$ from the above equation then completes the
computation of the source term for the Poission equation
\begin{eqnarray}\label{eq:poisson}
\vec{\nabla}^{2}\Phi =
{{\kappa}\over{2}}\left[C(\varphi)\rho_{\rm{DM}} -
C(\bar{\varphi})\bar{\rho}_{\rm{DM}} + \delta\rho_{\rm{B}}
- 2\delta V(\varphi)\right], 
\end{eqnarray}
where
$\delta\rho_{\rm{B}}\equiv\rho_{\rm{B}}-\bar{\rho}_{\rm{B}}$
and $\delta V(\varphi)\equiv V(\varphi)-V(\bar{\varphi})$ are
respectively the density perturbations of baryons and scalar field
(we have neglected perturbations in the kinetic energy
of the scalar field because it is always very small for our
model).

\subsection{Specification of Model}

\label{subsect:specification}

As mentioned above, to fully fix a model we need to specify the
functional forms of $V(\varphi)$ and $C(\varphi)$.
Here we will use the models investigated by
\cite{lz2009, lz2010, li2011}, with
\begin{eqnarray}\label{eq:coupling}
C(\varphi) = \exp(\gamma\sqrt{\kappa}\varphi),
\end{eqnarray}
and 
\begin{eqnarray}\label{eq:pot_chameleon}
V(\varphi) = {{\Lambda}\over{\left[1-\exp\left(-\sqrt{\kappa}\varphi\right)\right]^\alpha}}.
\end{eqnarray}
In the above $\Lambda$ is a parameter of mass dimension four and 
is of order the present dark energy density 
($\varphi$ plays the role of dark energy in the models). $\gamma, \alpha$ are dimensionless  parameters controlling the strength of the 
coupling and the steepness of the potentials respectively. 

We shall choose $\alpha\ll1$ and $\gamma>0$ as in \citet{lz2009, lz2010},
ensuring that $V_{eff}$ has a global minimum close to $\varphi=0$
and $d^2V_{eff}(\varphi)/d\varphi^2 \equiv m^2_{\varphi}$ at this
minimum is very large in high density regions. There are two
consequences of these choices of model parameters: (1) $\varphi$ is
trapped close to zero throughout  cosmic history so that
$V(\varphi)\sim\Lambda$ behaves as a cosmological constant; 
(2) the fifth force is strongly
suppressed in high density regions where $\varphi$ acquires a large
mass, $m^2_{\varphi}\gg H^2$ ($H$ being the Hubble expansion rate),
and thus the fifth force cannot propagate far. The suppression of the
fifth force is even stonger at early times, thus its influence on
structure formation occurs mainly at late times.
The environment-dependent behaviour of the scalar field was
first investigated by \citet{kw,ms}, and is often referred
to as the `chameleon effect'.

\subsection{Solutions in Spherical Symmetric Systems}

\label{subsect:spherical}

In this subsection we summarise the solutions to the radial profile of
the scalar field $\varphi$ in a spherically symmetric top-hat
overdensity with radius $R_{\rm TH}$, and (constant) matter density
$\rho_{\rm in}$ ($\rho_{\rm out}$) inside (outside) $R_{\rm TH}$. Such
a spherically symmetric system will be used to model dark matter halos
later. More details concerning these solutions can be found
in \cite{kw}.

If $\rho_{\rm in}=\rho_{\rm out}$, namely the mater density
is the same everywhere, then $\varphi$ will be constant across the
whole space and its value simply minimises the effective potential
$V_{eff}$. When $\rho_{\rm in}\neq\rho_{\rm out}$, $V_{eff}$ is
minimised by $\varphi_{\rm in}$ and $\varphi_{\rm out}$ inside and
outside $R_{\rm TH}$ respectively, while $\varphi$ will develop a
non-trivial radial profile.

Suppose we go towards the centre of the sphere from outside. If the
difference between $\varphi_{\rm in}$ and $\varphi_{\rm out}$ is
small, then $\varphi$ will settle to $\varphi_{\rm in}$ (from
$\varphi\sim\varphi_{\rm out}$ outside) soon after we enter the
sphere; if, on the other hand, the difference is large, then 
 $\varphi$ may never settle to $\varphi_{\rm
  in}$ even at the centre of the sphere. \cite{kw} give an estimate of
the distance $\Delta R$ that is needed for $\varphi$ to settle to
$\varphi_{\rm in}$ from $R_{\rm TH}$:
\begin{eqnarray}\label{eq:thinshell1}
\frac{\Delta R}{R_{\rm TH}} &=& \frac{\sqrt{\kappa}\varphi_{\rm out}-\sqrt{\kappa}\varphi_{\rm in}}{6\gamma\Phi_{\rm TH}},
\end{eqnarray} 
where $\Phi_{\rm TH}$ is the Newtonian potential at the surface of the sphere:
\begin{eqnarray}
\Phi_{\rm TH} \ =\ \frac{\kappa}{8\pi}\frac{M_{\rm TH}}{R_{\rm TH}}\ =\ \frac{\kappa}{6}\rho_{\rm in}R^2_{\rm TH},
\end{eqnarray}
and  $M_{\rm TH}\equiv\frac{4}{3}\pi R^3_{\rm TH}\rho_{\rm in}$ is the 
mass enclosed within the  sphere. Using this,
Eq.~(\ref{eq:thinshell1}) can be re-expressed as
\begin{eqnarray}\label{eq:thinshell2}
\frac{\Delta R}{R_{\rm TH}} &=& \frac{\varphi_{\rm out}-\varphi_{\rm in}}{\gamma\sqrt{\kappa}\rho_{\rm in}R^2_{\rm TH}}.
\end{eqnarray}

\cite{kw} present the solutions to $\varphi$ in two regimes. In the thin-shell regime, where $\Delta R\ll R_{\rm TH}$, the
solution is approximately
\begin{eqnarray}\label{eq:profile_thin}
&&\varphi(r)\nonumber\\ 
&=& \left\{%
\begin{array}{ll}
\varphi_{\rm in}, & \hbox{$r\in[0,R_0]$;} \\
\varphi_{\rm in}+\frac{\sqrt{\kappa}\gamma}{3}\rho_{\rm in}\left[\frac{r^2}{2}+\frac{R^3_{0}}{r}-\frac{3}{2}R^2_0\right], & \hbox{$r\in[R_0,R_{\rm TH}]$.}\ \ \ \\
\varphi_{\rm out}-\frac{\Delta R}{R_{\rm TH}}\frac{\sqrt{\kappa}\gamma\rho_{\rm in}R^3_{\rm TH}}{r}e^{-m_{\rm out}\left(r-R_{\rm TH}\right)}, & \hbox{$r\in[R_{\rm TH},\infty]$;} \\
\end{array}%
\right.
\end{eqnarray}
in which $R_0\in\left(0,R_{\rm TH}\right)$ and $R_{\rm TH}-R_0\ll R_{\rm TH}$; $m_{\rm out}$ is the effective mass of the scalar field
outside the sphere, which is given by
\begin{eqnarray}
m_{\rm out}^2 &\equiv& \frac{d^2V_{eff}\left(\varphi_{\rm out}\right)}{d\varphi^2}.
\end{eqnarray}
In the thick-shell regime, where $\Delta R>R_{\rm TH}$, the solution is approximately
\begin{eqnarray}\label{eq:profile_thick}
&&\varphi(r)\nonumber\\ 
&=& \left\{%
\begin{array}{ll}
\varphi_{\rm out}-\frac{3}{\sqrt{\kappa}}\gamma\Phi_{\rm TH}+\frac{1}{6}\gamma\sqrt{\kappa}\rho_{\rm in}r^2, & \hbox{$r\in[0,R_{\rm TH}]$;} \\
\varphi_{\rm out}-\frac{\sqrt{\kappa}\gamma\rho_{\rm in}R^3_{\rm TH}}{3r}e^{-m_{\rm out}\left(r-R_{\rm TH}\right)}, & \hbox{$r\in[R_{\rm TH},\infty]$.}\ \ \\
\end{array}%
\right.
\end{eqnarray}
Physically, if $\varphi$ has developed a thin shell near the edge of
the spherical overdensity, then from Eq.~(\ref{eq:profile_thin}) we
can see that only a fraction $\Delta R/R_{\rm TH}$ of the total mass
enclosed in $R_{\rm TH}$ contributes to the fifth force on a test
particle at the edge.  This means that the fifth force from the matter
inside the sphere is strongly screened.  In the thick-shell regime the
fifth force is not screened.

Note that in the thick shell regime at the edge of the halo we have
\begin{eqnarray}
\frac{C_\varphi}{C}\nabla\varphi = \gamma\frac{d}{dr}\left[\sqrt{\kappa}\varphi(r)\right] = -2\gamma^2\frac{d\Phi_{\rm TH}}{dr},
\end{eqnarray}
which indicates that the magnitude of the fifth force is $2\gamma^2$ times that of gravity, and its effect is to rescale the Newton's constant by $1+2\gamma^2$.

\section{Analytical Method for Structure Formation}

\label{sect:method}

Having reviewed the chameleon model and the solutions in spherically
symmetric top-hat overdensities, let us now turn to excursion set
theory \citep{bcek}, which was  developed to study structure
formation in cold dark matter scenarios. We will generalise
the excursion set approach to the
chameleon model, where the dark matter particles experience an extra,
environment-dependent, fifth force.

\subsection{Excursion Set Theory}

\label{subsect:excursion}

It is widely accepted that the large-scale structure (LSS) in the
Universe has developed hierarchically through gravitational
instability.  The excursion sets (regions where the matter
density exceeds some threshold when filtered on a suitable scale)
generally correspond to sites of formation of virialised structures
\citep{ss1988,ck1988,ck1989,efwd1988,er1988,nw1987,cc1988}.

The filtered, or smoothed, matter density perturbation field
$\delta({\bf x}, R)$, is given by
\begin{eqnarray}\label{eq:smooth}
\delta({\bf x}, R) &=& \int W(|{\bf x}-{\bf y}|; R)\delta({\bf y})d^3{\bf y}, \nonumber\\
&=& \int \tilde{W}(k; R)\delta_{\bf k}e^{i\bf{k\cdot x}}d^3{\bf k},
\end{eqnarray}
where  $W({\bf r}; R)$ is a filter, or window function,
 with radius $R$, and $\tilde{W}(k;R)$ its Fourier transform; $\delta({\bf x})\equiv\rho({\bf x})/\bar{\rho}-1$ is the 
true, unsmoothed, density perturbation field and $\delta_{\bf k}$ its Fourier transform; we will always use an overbar to denote background quantities.

As usual, we assume that the initial density perturbation field
$\delta({\bf x})$ is Gaussian and specified by its power spectrum
$P(k)$. The root-mean-squared (rms) fluctuation of mass in the
smoothing window is given by
\begin{eqnarray}\label{eq:rms_fluc}
S(R)\ \equiv\ \sigma^2(R)\ \equiv\ \langle\delta^2({\bf x}; R)\rangle\ =\ \int P(k)\tilde{W}(k; R)d^3{\bf k}.
\end{eqnarray}
 Note that, given the power spectrum $P(k)$, $S$, $R$ and $M$ are
 equivalent measures of the scale of a spherical perturbation and they
 will be used interchangeablly below.

If $\tilde{W}(k;R)$ is chosen to be a sharp filter in k-space, then
the increment of $\delta({\bf x};R)$ as $R\rightarrow R-\delta R$
or equivalently $S\rightarrow S+\delta S$ comes from only the extra
higher-$k$ modes of the density perturbation (see
Eq.~(\ref{eq:smooth})). The absence of correlation between these
different wavenumbers 
means that the increment of $\delta({\bf x};R)$ is independent of its
previous value (the Markov property). It is also a Gaussian field,
with zero mean and variance $\delta S$. Thus, considering $S$ as a
`time' variable, we find that $\delta({\bf x};S)$ can be described by
a Brownian motion.

The probability distribution of $\delta({\bf x};R)$ is a Gaussian
\begin{eqnarray}\label{eq:gaussian}
P(\delta,S)d\delta = \frac{1}{\sqrt{2\pi S}}\exp\left[-\frac{\delta^2}{2S}\right]d\delta.
\end{eqnarray}
In an Einstein-de Sitter or a $\Lambda$CDM universe, the linear growth
of initial density perturbations is scale-independent, so that
$\delta({\bf x})$ and $\sigma(R)=\sqrt{S}$ grow in the same manner,
and as a result the density field will remain Gaussian while
it is linear. Following the
standard literature, hereafter we shall use $\delta({\bf x}; R)$ to
denote the initial smoothed density perturbation extrapolated to the
present time using linear perturbation theory, and the same for
$\sigma$ or $S$.

In the standard cold dark matter scenario, the initial smoothed
densities which, extrapolated to the present time, equal (exceed)
$\delta_c$ correspond to regions where virialised dark matter halos
have formed today (earlier). In an Einstein-de Sitter universe
$\delta_c$ is a constant, while in a $\Lambda$CDM universe it depends
on the matter density $\Omega_m$. In neither case does $\delta_c$
depend on the size of (or equivalently the mass enclosed in) the
smoothed overdensity, or the environment surrounding the overdensity.

As a result, to see if a spherical region with initial radius $R$ has
collapsed to virialised objects today or lives in some larger region
which has collapsed earlier, we only need to see whether $\delta({\bf
  x}; \geq R)\geq\delta_c$. Put another way, the fraction of the
total mass that is incorporated in virialised dark matter halos
heavier than $M=\frac{4}{3}\pi R^3\bar{\rho}_i$ is just the fraction
of the Brownian motion trajectories $\delta({\bf x}; S)$ which have
crossed the constant barrier $\delta_c$ by the 'time' $S=S(R)$, which
is given by \citep{bcek}
\begin{eqnarray}
F(M,z) = \frac{1}{\sqrt{2\pi S}}\int^\infty_{\frac{D_+(0)}{D_+(z)}\delta_c}\left[e^{-\frac{\delta^2}{2S}}-e^{-\frac{(\delta-2\delta_c)^2}{2S}}\right]d\delta,
\end{eqnarray} 
where the lower limit of the integral is $\frac{D_+(0)}{D_+(z)}\delta_c$, because if a virialised object formed at redshift $z$, then its corresponding initial smoothed density linearly extrapolated to $z$ is 
$\delta_c$, while extrapolated to today it is $\frac{D_+(0)}{D_+(z)}\delta_c$ with $D_+(z)$ being the linear growth factor at $z$. In Einstein-de Sitter cosmology $D_+(z)\propto(1+z)^{-1}$ and
this quantity becomes $(1+z)\delta_c$.

Alternatively, one can say that the fraction of the total mass that is
incorporated in halos,  the radii of which fall in $[R,R+\delta R]$ (or
equally $[S,S+\delta S]$) and which collapse at $z=z_f$ is given
by
\begin{eqnarray}
f(S,z_f)dS = \frac{1}{\sqrt{2\pi S}}\frac{D_+(0)\delta_c}{D_+(z_f)S}\exp\left[-\frac{D_+^2(0)\delta_c^2}{2D_+^2(z_f)S}\right]dS,
\end{eqnarray}
where $f(S)$ the distribution of the first-crossing time of the Brownian motion to the barrier $D_+(z=0)\delta_c/D_+(z=z_f)$. Once this is obtained, one can compute the halo mass function observed at $z_f$ as
\begin{eqnarray}
\frac{dn(M)}{dM}dM = \frac{\bar{\rho}_m(z_f)}{M}f(S)dS.
\end{eqnarray}
Other observables, such as the dark matter halo bias \citep{mw1996},
merger history \citep{lc1993}, void distribution \citep{sv2004}, can
be computed with certain straightforward generalisations of the theory.

\subsection{Characterising the Chameleon Effect}

\label{subsect:chameleon}

To incorporate the chameleon effect into the model, we need to have
some idea about which physical quantities are most relevant 
and how they might affect the analysis. In our study of dark halo
formation based on spherical collapse of top-hat overdensities,
Eqs.~(\ref{eq:thinshell2}, \ref{eq:profile_thin},
\ref{eq:profile_thick}) roughly characterise where the chameleon
effect is strong using the following relevant physical quantities (in
addition to the parameters $\alpha$ and $\gamma$ which are fixed once
a model is specified):
\begin{enumerate}
\item $\varphi_{\rm out}$, the value of $\varphi$ which minimises
  $V_{eff}(\varphi)$ outside the sphere. This in turn depends on the
  matter density $\rho_{\rm out}$ which we take approximately by
  smoothing the density field using a filter centred at our sphere
  with a radius \tcr{$\xi$}. Evidently, $\rho_{\rm out}$
  describes the environment-dependence of the chameleon effect, \tcr{while $\xi$ is the size of the environment, which itself is modelled as a spherical top-hat overdensity or underdensity.} 
\item $\varphi_{\rm in}$, that minimises $V_{eff}(\varphi)$ inside the spherical halo. This depends on $\rho_{\rm in}$, which is the density of the spherical halo. 
\item $R_{\rm TH}$, the radius of the top-hat spherical halo.
\end{enumerate} 
In summary, there are three quantities which determine the strength of
the chameleon effect: $\rho_{\rm out}, \rho_{\rm
  in}$ and $R_{\rm TH}$, of which the latter two characterise the
spherical halo under study while the former represents the local
environment in which the halo is located.

The complexity, however, is that all of these three quantities evolve
in time, and they can all be different for different halos. In
particular, $\rho_{\rm in}, \rho_{\rm out}$ are the true {\it
  non-linear densities} inside and outside the halo {\it at arbitrary
  redshifts $z\geq0$}, while in the excursion set approach we are
dealing with {\it overdensities} which are {\it linearly extrapolated
to the present day}. We must be able to relate the former to
the latter to facilitate a statistical treatment based on the Gaussian
distribution of the linear matter density perturbation field.

\subsection{Fixed-scale Environment Approximation}

\label{subsect:fixed-scale}

In considering the linearly extrapolated matter density
field, we must decide whether the linear evolution should be
computed as in  $\Lambda$CDM or the chameleon model?
Since we assume that the chameleon model starts with the same
initial conditions as the $\Lambda$CDM model, and the linear
perturbation for the latter is much easier to compute, in what
follows we shall always use the $\Lambda$CDM-linearly-extrapolated
$\delta({\bf x}; R)$.

Let us consider the non-linear evolution of a smoothed density
perturbation $\delta({\bf x};R)$ which is surrounded by another
top-hat sphere with $\Lambda$CDM-extrapolated density perturbation
\tcr{$\delta_{\rm env}({\bf x};\xi)$. It is evident that to specify
  the environment we need to know the value of $\xi$.}

\tcr{There are certain guidelines in the choice of $\xi$. To represent the
local environment, $\xi$ can not be too large because otherwise the
matter density within $\xi$ would simply be the background
value $\bar{\rho}_m$. $\xi$ cannot be too small either, because
the environment should be significantly larger than the hosted dark
matter halo to be compatible to the characteristic length scale 
on which the scalar field value changes 
from $\varphi_{\rm in}$ to $\varphi_{\rm out}$. These considerations
suggest that the natural choice of $\xi$ is a few times the
virial radius of the hosted halo. However, this means that $\xi$ is
dependent on both time and halo size, precluding a simple
analytic extension of the excursion set approach.} 

\tcr{Since we are interested in this paper in qualitative (rather than
high precision) results, we adopt a  {\it fixed-scale
environment approximation}, in which $\xi$ is taken
to be a  constant. 
As a simple choice, we adopt $\xi=8h^{-1}$~Mpc, where
$h=H_0/100~{\rm km/s/Mpc}$ and $H_0$ is the present Hubble
constant. As shown in Fig.~4 of \citet{lz2010}, the length scale of
the spatial variation of the scalar field value ($\varphi_{\rm
out}$) is typically a few Mpc at late times, which is roughly the
same as $\xi$. Such a large scale is well beyond the Compton length
of the scalar field $\varphi$, and so the fifth force is not
expected to play an important role. As the cosmic background
expansion in the chameleon model is indistinguishable from that of
$\Lambda$CDM as well, the non-linear evolution of the spherical
overdensity enclosed by $\xi$ is well described by
$\Lambda$CDM. This means that we can relate $\delta_{\rm env}({\bf
x};\xi)$ to $\Delta_{\rm env}({\bf x};\xi)$ (we shall use $\Delta$
to represent non-linear density contrasts throughout this paper)
using the $\Lambda$CDM spherical collapse model, and then $\rho_{\rm
out}=\bar{\rho}\left[1+\Delta_{\rm env}({\bf x};\xi)\right]$. In
this way, we have related $\rho_{\rm out}$ to $\delta_{\rm env}({\bf
 x};\xi)$.}


Assuming no shell crossing, the mass enclosed by the (comoving)
smoothing radius $R$ is $M=\frac{4}{3}\pi\bar{\rho}R^3$. With
$\rho_{\rm out}$ at arbitrary time known, we can calculate the
  evolution of the initial density perturbation corresponding to
  $\delta({\bf x};R)$ since (i) we know the strength of the fifth
  force at arbitrary time from
  Eqs.~(\ref{eq:profile_thin},\ref{eq:profile_thick}), and (ii) we can
  compute the collapse history of the sphere, namely $R_{\rm TH}$:
  because of mass conservation, $\frac{4}{3}\pi\rho_{\rm in}R^3_{\rm
    TH}=M$, giving $\rho_{\rm in}$ in terms of $M$ (equivalently $R$)
  and $R_{\rm TH}$, and this can be used to quantify the
  chameleon effect for the next step.
  
  As a result, once a top-hat overdensity $\delta({\bf x};R)$ and its
environment $\delta_{\rm env}({\bf x};\xi)$ are fixed, we can
determine its collapse history.

\subsection{Spherical Collapse}

\label{subsect:collapse}

We have seen above that the spherical collapse of a top-hat
overdensity is specified by $\delta({\bf x};R)$ and $\delta_{\rm
  env}({\bf x};\xi)$. Now we shall use these quantities to calculate
the critical ($\Lambda$CDM-linearly-extrapolated) density contrast
$\delta_c({\bf x}; R,z_f,\delta_{\rm env}({\bf x};\xi))$ that is
needed for an initial overdensity with radius $R$, residing in
environment $\delta_{\rm env}({\bf x};\xi)$, to collaspe into a
virialised object at redshift $z_f$ in the chameleon model. In
the Einstein-de Sitter and $\Lambda$CDM cosmologies $\delta_c$ does
not depend on $R$ or $\delta_{\rm env}$, but in the chameleon model
these quantities are crucial in determining the effect of the fifth
force.

In the chameleon models considered here, the choice of parameters $\alpha$ and $\gamma$, as mentioned above, ensures that the background cosmic expansion is practically indistinguishable from that of $\Lambda$CDM \citep{lz2009}. For simplicity, the evolution of the scale factor $a(t)$ is specified as
\begin{eqnarray}\label{eq:friedman}
\frac{H^2}{H_0^2} &=& \Omega_m a^{-3} + \Omega_\Lambda,
\end{eqnarray}
with $H\equiv\dot{a}/a$ and the overdot denotes the (physical) time
derivative. Throughout this paper we shall adopt $\Omega_m=0.24$,
$\Omega_\Lambda=0.76$ and $H_0=71.9$~km/s/Mpc. Also note that our
study is limited to late times, when structure becomes non-linear,
which is why radiation is not included in this and subsequent
equations.

\subsubsection{Evolution of Overdensities in the $\Lambda$CDM Model}

Let us consider first the linear and non-linear evolution for an initial density perturbation in the $\Lambda$CDM model, which will be used to calculate the relation between $\delta_{\rm env}({\bf x};\xi)$ and $\Delta_{\rm env}({\bf x},a;\xi)$ (here we have written explicitly the dependence of $\Delta_{\rm env}$ on time or equivalently $a$ or $z$). The convention and definitions here follow closely that of, e.g., \cite{valageas2009}.

The linear evolution of the density perturbation satisfies,
\begin{eqnarray}
\ddot{\delta}+2H\dot{\delta}-\frac{1}{2}\kappa\bar{\rho}_m\delta &=& 0.
\end{eqnarray}
Using equations (24) and (25), 
it is straightforward to show that the linear growth factor $D_+$ satisfies 
\begin{eqnarray}\label{eq:dplus}
D''_++\left[2-\frac{3}{2}\Omega_m(N)\right]D'_+-\frac{3}{2}\Omega_m(N)D_+ &=& 0,
\end{eqnarray}
in which a prime means derivative with respect to $N\equiv\ln a$, and
\begin{eqnarray}
\Omega_m(N) &\equiv& \frac{\Omega_me^{-3N}}{\Omega_me^{-3N}+\Omega_\Lambda},\\
\Omega_\Lambda(N) &\equiv& \frac{\Omega_\Lambda}{\Omega_me^{-3N}+\Omega_\Lambda},
\end{eqnarray}
are respectively the fractional densities for matter and dark energy at arbitrary $N$. The initial conditions are given by the fact that, deep into the matter dominated era, $D_+(a_i)=a_i$, and therefore $D'_+(a_i)=a_i$.

To analyse non-linear spherical collapse, let us denote the physical
radius of the considered spherical halo at time $t$ by $r(t)$, and its
physical radius if it has  not collapsed by $q(t)=a(t)R$
(remember that $R$ is the comoving radius of the filter). Because of
the spherical symmetry, it is straightforward to write down the
evolution equation for $r(t)$ as
\begin{eqnarray}\label{eq:revolution}
\frac{\ddot{r}}{r} &=& -\frac{\kappa}{6}\left(\rho_m-2\rho_\Lambda\right),
\end{eqnarray}
where $\rho_m\equiv3M/4\pi r^3$ is the true matter density in the halo and the constant $\rho_\Lambda$ is the dark energy density. Let us define $y(t)=r(t)/q(t)$ and change the time variable to $N$. By using Eqs.~(\ref{eq:friedman}, \ref{eq:revolution}) and $q(t)\propto a(t)$, it can be shown that 
\begin{eqnarray}\label{eq:yevolution}
y'' + \left[2-\frac{3}{2}\Omega_m(N)\right]y' + \frac{\Omega_m(N)}{2}\left(y^{-3}-1\right)y &=& 0,
\end{eqnarray}
which is clearly a non-linear equation. At very early times we must have $y\approx1$ and we can write $y=1+\epsilon$ with $|\epsilon|\ll1$. Substituting this into Eq.~(\ref{eq:yevolution}) to get the linearised evolution equation for $\epsilon$, and comparing with Eq.~(\ref{eq:dplus}), we find that $\epsilon\propto D_+$, in which the proportional coefficient could be found using mass conservation $y^3(1+\delta_i)=1\Rightarrow \epsilon=-\delta_i/3\propto D_+$ (here $\delta_i$ is the linear density perturbation at the initial time). As a result, the initial conditions for $y$ are $y(a_i)=1-\delta_i/3$ and $y'(a_i)=-\delta_i/3$.

Eqs.~(\ref{eq:dplus}, \ref{eq:yevolution}), associated with their corresponding initial conditions, completely determine the necessary dynamics in the $\Lambda$CDM model.

\subsubsection{Evolution of Overdensities in the Chameleon Model}

With the preliminaries given above, we can now consider spherical
collapse in the chameleon model. 

 From the discussions in
\S~\ref{subsect:spherical} and results of \cite{lz2009}, we know that
the fifth force acts as if it renormalises the Newton's constant by
$(1+2\gamma^2)$ if the chameleon effect is weak ({\it i.e.}, in the
thick-shell regime); on the other hand, it is strongly suppressed in
the thin-shell regime. In particular, comparison of
Eqs.~(\ref{eq:profile_thin}, \ref{eq:profile_thick}) shows that the
two regimes give the same exterior solution when $\Delta R/R_{\rm
  TH}=1/3$. Therefore, we propose to approximately take account of the
effect of the fifth force as if it effectively rescales the Newton's
constant by $1+2\gamma^2\min\left\{3\Delta R/R_{\rm TH},
1\right\}$. This is certainly not expected to be very accurate, but
our aim here is to present a method which captures the essential
features of the environment dependence.

Because we do not need the the linear perturbation evolution in the
chameleon model, we shall go to the spherical collapse
directly. According to the above approximation, the equation of motion
of a spherical shell at the edge of the top-hat overdensity is
\begin{eqnarray}\label{eq:revolution2}
\frac{\ddot{r}}{r} &=& \frac{1}{3}\kappa\rho_\Lambda - \frac{1}{6}\kappa\rho_m\left[1+2\gamma^2\min\left\{\frac{3\Delta R}{R_{\rm TH}},1\right\}\right],
\end{eqnarray}
where we have neglected the perturbation in the energy density of the
scalar field and its kinetic energy, which are negligible
\citep{lz2009}. Note that this means that the energy density of the
scalar field is the same as that of the vacuum energy in $\Lambda$CDM
model.

The scalar field value which minimises the effective potential $V_{eff}(\varphi)$ is given by \citep{lz2009}
\begin{eqnarray}
\sqrt{\kappa}\varphi &\approx& \frac{\alpha}{\gamma}\frac{V_0}{\rho_m},
\end{eqnarray}
where $\rho_m$ is the local matter density. Substituting this into Eq.~(\ref{eq:thinshell1}), we find that, at time $a$,
\begin{eqnarray}\label{eq:shelly}
\frac{3\Delta R}{R_{\rm TH}} &\approx& \frac{1}{\left(H_0R\right)^2}\frac{\alpha}{\gamma^2}\frac{\Omega_\Lambda}{\Omega_m^2}\left[\frac{y^3_{\rm env}}{1+\delta_{{\rm env},i}}-\frac{y^3_{\rm h}}{1+\delta_{i}}\right]y_{\rm h}a^4,
\end{eqnarray}
in which $y_{\rm h}$ is the $y$ for the considered halo and $y_{\rm env}$ that for the environmental spherical overdensity smoothed at radius $\xi$. $\delta_{{\rm env},i}$ and $\delta_{i}$ are respectively the initial values for $\delta_{\rm env}({{\bf x};\xi})$ and $\delta({{\bf x};R})$ and
\begin{eqnarray}\label{eq:assis}
\delta_{{\rm env},i} &=& \frac{D_+(z=z_i)}{D_+(z=0)}\delta_{\rm env}({\bf x};\xi),\nonumber\\
\delta_i &=& \frac{D_+(z=z_i)}{D_+(z=0)}\delta({\bf x};R).
\end{eqnarray}
In the derivation of Eq.~(\ref{eq:shelly}) we have used the approximation that masses are conserved within the top-hat overdensities with radii $R$ and $\xi$. Note that because of the unit convention $c=1$ the quantity $H_0R$ is dimensionless. Eq.~(\ref{eq:shelly}) shows that the effects of the fifth force will be more suppressed by:
\begin{enumerate}
\item increasing $\gamma$ and decreasing $\alpha$, both making the scalar field heavier and unable to propagate far;
\item increasing $\Omega_m$, meaning that the matter density 
is higher in the Universe, again making the scalar field heavier;
\item increasing environmental density $\delta_{\rm env}({\bf
  x};\xi)$, therefore making the term in the brackets smaller;
\item considering earlier times, where $a$ is smaller, because the
  overall matter density is higher then, and
\item considering bigger halos (larger $R$), which are more efficient
  in screening the fifth force.
\end{enumerate} 

From the earlier discussion, $y_{\rm env}$ is governed by Eq.~(\ref{eq:yevolution}), and now we need to find an evolution equation for $y_{\rm h}$ as well. This can be obtained from Eq.~(\ref{eq:revolution2}) following
the derivation of  Eq.~(\ref{eq:yevolution}). The result is
\begin{eqnarray}\label{eq:yevolution2}
&&y_{\rm h}'' + \left[2-\frac{3}{2}\Omega_m(N)\right]y_{\rm h}'\nonumber\\ 
&=& -\frac{\Omega_m(N)}{2}\left(y_{\rm h}^{-3}-1\right)y_{\rm h}\left[1+2\gamma^2\min\left\{\frac{3\Delta R}{R_{\rm TH}},1\right\}\right],
\end{eqnarray}
where $3\Delta R/R_{\rm TH}$ is given by
Eq.~(\ref{eq:shelly}). Because at very early times the chameleon
effect is very strong, the initial conditions of this equation can be
chosen exactly as in the $\Lambda$CDM model. Eqs.~(\ref{eq:dplus},
\ref{eq:yevolution}, \ref{eq:yevolution2}), together with
Eqs.~(\ref{eq:shelly}, \ref{eq:assis}) form a closed system for our
chameleon model.These  completely fix the evolution
of a spherical overdensity $\delta({\bf x};R)$ residing in the
environment $\delta_{\rm env}({\bf x};\xi)$. Note that
Eq.~(\ref{eq:dplus}) only needs to be solved once.

\begin{figure}
\includegraphics[width=88mm]{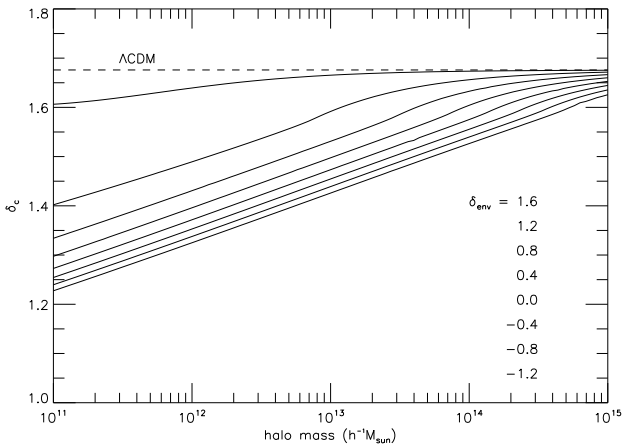}
\caption{The critical ($\Lambda$CDM-linearly-extrapolated) density perturbation $\delta_c$ for the given spherical overdensity with mass $M$ to collapse at $z_f=0$. Shown are $\delta_c$ as functions of $M$ for halos residing in different environments, with (solid curves from top to bottom) $\delta_{\rm env}=1.6, 1.2, 0.8, 0.4, 0, -0.4, -0.8, -1.2$. For comparison, the constant $\delta_c\approx1.676$ for the $\Lambda$CDM model is overplotted as the dashed horizontal line.}
\label{fig:dc_env}
\end{figure}

\begin{figure}
\includegraphics[width=88mm]{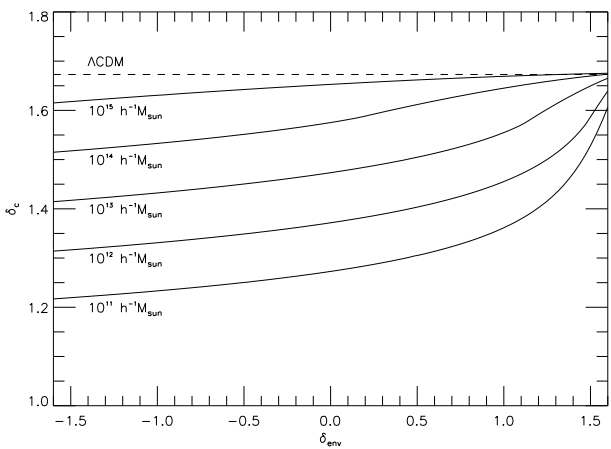}
\caption{The critical ($\Lambda$CDM-linearly-extrapolated) density perturbation $\delta_c$ for the given spherical overdensity residing in environment $\delta_{\rm env}$ to collapse at $z_f=0$. Shown are $\delta_c$ as functions of $\delta_{\rm env}$ for halos with different masses (solid curves from top to bottom) $M=10^{15}, 10^{14}, 10^{13}, 10^{12}, 10^{11}~h^{-1}M_{\rm sun}$. For comparison, the constant $\delta_c\approx1.676$ for the $\Lambda$CDM model is overplotted as the dashed horizontal line.}
\label{fig:dc_mass}
\end{figure}

\begin{figure}
\includegraphics[width=88mm]{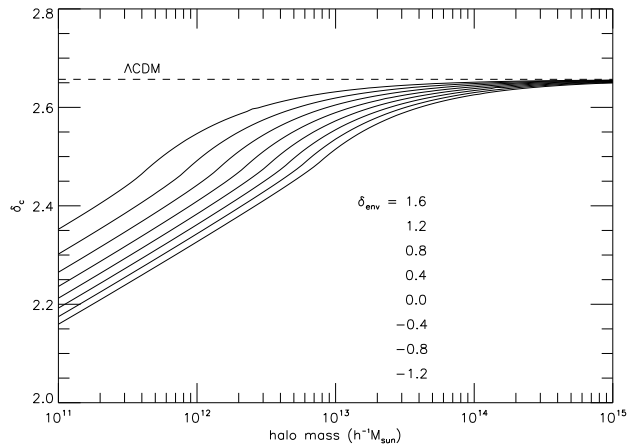}
\caption{Same as Fig.~\ref{fig:dc_env} but for spherical overdensities which collapse at $z_f=1$. In this case $\delta_c\approx2.657$ for the $\Lambda$CDM model (the horizontal dashed line).}
\label{fig:dc_envb}
\end{figure}

%
%

\subsubsection{Numerical Examples}

To get an idea about how the environment-dependent fifth force changes
 spherical collapse in the chameleon model, we present some
numerical examples in this section.

As we have discussed above, the critical density (linearly extrapolated to today using $\Lambda$CDM model) which is needed for a spherical overdensity to collapse at redshift $z_f$ depends on $R$ (the spherical overdensity's own property) and $\delta_{\rm env}$ (its environment): $\delta_c({\bf x})=\delta_c({\bf x};R,\delta_{\rm env},z_f)=\delta_c({\bf x};M,\delta_{\rm env},z_f)$ where we have used $M\approx\frac{4}{3}\pi\bar{\rho}_{m0}R^3$. 

Fig.~\ref{fig:dc_env} shows $\delta_c$ as a function of halo mass $M$ for different values of $\delta_{\rm env}$ and $z_f=0$. We have considered halos in eight different environments with $\delta_{\rm env}$ ranging between $1.6$ (very dense environment) and $-1.2$ (very empty envrionment). As can be seen there, the fifth force lowers $\delta_c$ compared with the $\Lambda$CDM result (dashed line), which is as expected because it makes collapse easier. Note that
\begin{enumerate}
\item Unlike in $\Lambda$CDM, in the chameleon model $\delta_c$ is mass and therefore scale dependent, a point which we will return to later.
\item For a given $\delta_{\rm env}$, $\delta_c$ is closer to the $\Lambda$CDM result for more massive halos because these halos are more efficient in screening the fifth force [see also Eq.~(\ref{eq:thinshell2})]. Note however that $\delta_c$ will never exceed the corresponding value in $\Lambda$CDM model because the fifth force always helps rather than prevents the collapse.
\item For a given halo mass $M$, $\delta_c$ is closer to the $\Lambda$CDM prediction in denser environments, where the chameleon effect is stronger.
\end{enumerate}
These can also be seen in Fig.~\ref{fig:dc_mass}, which shows $\delta_c$ as a function of $\delta_{\rm env}$ for different halo masses.

Fig.~\ref{fig:dc_envb} shows the same results as Fig.~\ref{fig:dc_env}, but for the halos which collapse at $z_f=1$. This shows similar qualitative behaviour as does the $z_f=0$ case, but the relative difference between the collapsing threshold $\delta_c$ and its $\Lambda$CDM result is generally smaller because by $z=1$ the fifth force is strongly suppressed by the chameleon mechanism in most environments and because the halos which form at $z=0$ experience the fifth force for longer.

The simplified computation described in this section can capture the essential effects of the chameleon fifth force. We will use it as an ingredient of the extended excursion set model to be introduced below.

\subsubsection{Notes on the Approximations Used}

As mentioned earlier, the purpose of this work is to introduce 
a conceptually simple, largely analytic,
method of incorporating environment dependence in the study of
structure formation that is adequate for parameter exploration.
Consequently we have used a number of approximations to
simplify the calculation. Here we briefly summarise these
approximations and discuss how they can be improved using numerical
methods:
\begin{enumerate}
\item The computation of the scalar field profile $\varphi(r)$ in the spherical halo: in this work we have adopted the analytical approximations given in \cite{kw}, which could be improved by solving the scalar field EOM explicitly using numerical methods.
\item The detailed shape of the spherical halo: because of the
  environment dependence of the fifth force, shells at different radii
  of the halo will travel at different speeds, resulting in a
  modification to the top-hat shape of the halo. In this work we have
  assumed a constant overdensity for the halo, which is only an
  approximation. In general we expect that matter will 
  accumulate (slightly) towards the edge of the halo.
This effect can be computed
  accurately once $\varphi(r)$, or equivalently the profile of the
  fifth force is known precisely (see \cite{mss2009} for an example).
\end{enumerate}
We will leave these improvements to future work.

\subsection{Generalised Excursion Set Method for the Chameleon Model}

\label{subsect:generalisation}

We have seen above that the excursion set prediction of the halo mass
function (based on the spherical collapse model in the $\Lambda$CDM
cosmology) is closely related to the first crossing distribution of a
flat barrier by a Brownian random walk that starts from zero. In the
chameleon model two factors lead to a more complicated problem.
\begin{enumerate}
\item The barrier that is to be crossed by the Brownian motion is no longer flat, but rather depends on the mass scale $M$ (c.f.~Figs.~\ref{fig:dc_env} and \ref{fig:dc_envb}), or equivalently  $R$ or $S(R)$.
\item The barrier is also affected by the environment surrounding the
  collapsing halo (c.f.~Fig.~\ref{fig:dc_mass}), and so we
  need to know the probability distribution of its environment
  ($\delta_{\rm env}$) as well.
\end{enumerate}
These complications are the subject of  this section.

\subsubsection{Unconditional First Crossing of a Moving Barrier}

The distribution of the first crossing of a general barrier by a
Brownian motion has no closed-form analytical solutions except for
some simple barriers, {\it e.g.}, flat \citep{bcek} and linear
\citep{sheth1998,st2002}. Unfortunately  neither of these is
a good approximation to our general barrier
(cf.~Fig.~\ref{fig:dc_env}). As a result, we shall follow
\cite{zh2006} and numerically compute this distribution. We shall
briefly review their method for completeness.

Denote the unconditional probability that a Brownian motion starting
off at zero hits the barrier $\delta_c(S)$ for the first time in
$[S,S+dS]$ by $f(S)dS$. Then, $f(S)$, the probability density,
satisfies the following integral equation
\begin{eqnarray}
f(S) &=& g(S) + \int^S_0 dS'f(S')h(S,S'),
\end{eqnarray}
in which
\begin{eqnarray}
g(S) &\equiv& \left[\frac{\delta_c}{S}-2\frac{d\delta_c}{dS}\right]P\left(\delta_c,S\right),\nonumber\\
h(S,S') &\equiv& \left[2\frac{d\delta_c}{dS}-\frac{\delta_c-\delta'_c}{S-S'}\right]P(\delta_c-\delta'_c,S-S'),
\end{eqnarray}
where for brevity we have suppressed the $S$-dependence of $\delta_c(S)$ and used $\delta'_c\equiv\delta_c(S')$; $P(\delta, S)$ is given in Eq.~(\ref{eq:gaussian}). This equation could be solved numerically on an equally-spaced mesh on $S$: $S_i=i\Delta S$ with $i=0,1,\cdots,N$ and $\Delta S=S/N$. The solution is \citep{zh2006}
\begin{eqnarray}
f_0 &=& g_0 \ =\ 0,\nonumber\\
f_1 &=& (1-\Delta_{1,1})^{-1} g_1,\\
f_{i>1} &=& (1-\Delta_{1,1})^{-1}\left[g_i+\sum_{j=1}^{i-1}f_j(\Delta_{i,j}+\Delta_{i,j+1})\right],\nonumber
\end{eqnarray}
where we have used $f_i=f(S_i)$ and similarly for $g_i$ to lighten the notation, and defined 
\begin{eqnarray}
\Delta_{i,j} &\equiv& \frac{\Delta S}{2}h\left(S_i,S_j-\frac{\Delta S}{2}\right).
\end{eqnarray}
We have checked that this method agrees accurately with the
analytic solution for the 
flat-barrier crossing problem.

\begin{figure}
\includegraphics[width=88mm]{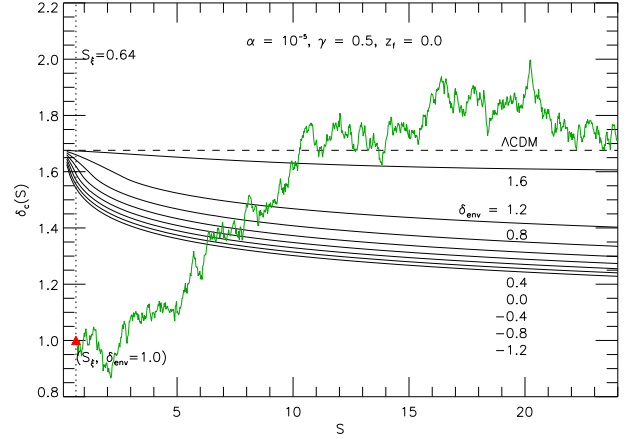}
\caption{(Colour online) The moving barriers $\delta_c(S)$ for different values of $\delta_{\rm env}$ as indicated beside the solid curves. The dashed line is the constant $\delta_c$ for spherical collapse in the $\Lambda$CDM model. The vertical dotted line represents $S=S_\xi=\sigma^2_8=0.64$, which gives the length scale used to define environment. Also plotted is the trajectory of a Brownian random walk which starts at $(S_\xi, \delta_{\rm env}=1.0)$ (the triangle). Note that the first crossing happens earlier in the chameleon models because the barrier is lower.}
\label{fig:bm}
\end{figure}

\subsubsection{Conditional First Crossing of a Moving Barrier}

The unconditional first crossing distribution, which relates directly to the halo mass function in the $\Lambda$CDM model, is not particularly useful in the chameleon model. This is because spherical overdensities in different environments will follow different evolution paths. If it is in the environment specified by $\left(\delta_{\rm env}, S_\xi\right)$, then $\left(\delta_{\rm env}, S_\xi\right)$ should be the starting point of the Brownian motion trajectory. In other words, we actually require the first crossing distribution {\it conditional on} the trajectory passing $\delta_{\rm env}$ at $S=S_\xi$. Note that in a broader sense the unconditional distribution is a conditional one with $\left(\delta_{\rm env}, S_\xi\right)=(0,0)$.

Evidently, $\delta_{\rm env}$ has its own distribution: very dense and very empty environments are both quite rare. To quantify this distribution, we need to first define the environment, or equally its smoothing scale $\xi$, which \tcr{has been chosen to be $8h^{-1}$Mpc above}. 

%

The problem then reduces to the calculation of the first crossing
probability conditional on the Brownian motion trajectory passing
$\delta_{\rm env}$ at $S_\xi=\sigma^2_8$: $f(\delta_c(S,\delta_{\rm
  env}),S~|~\delta_{\rm env},S_\xi)$, where we have written explicitly
the $\delta_{\rm env}$-dependence of $\delta_c$. The numerical
algorithm to calculate the conditional first crossing probability is a
simple generalisation of the one used above to compute the
unconditional first crossing probability \citep{phs2011} and is not
presented in detail here.

\begin{figure}
\includegraphics[width=88mm]{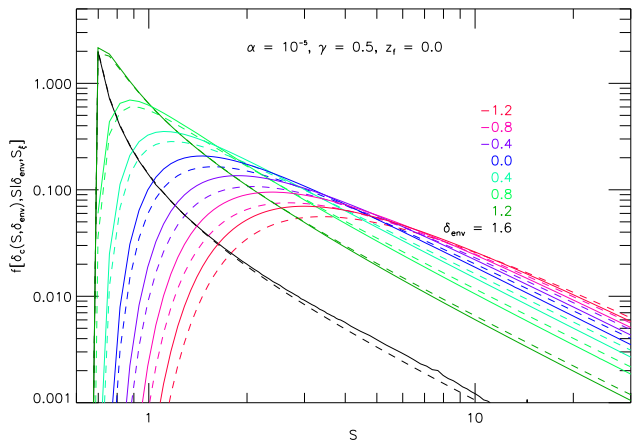}
\caption{(Colour online) The condition first-crossing distribution for Brownian random walks starting off at $(S_\xi, \delta_{\rm env})$ in the chameleon (the solid curves) and $\Lambda$CDM models (dashed curves). This depends sensitively on the values of $\delta_{\rm env}$ (indicated above the curves) as explained in the text. Some physical parameters are also shown.}
\label{fig:fS}
\end{figure}

Fig.~\ref{fig:bm} shows the moving barrier $\delta_c(S)$ as a function
of $S$ for different values of $\delta_{\rm env}$. As an illustration,
we have also shown a Brownian motion trajectory which passes
$\delta_{\rm env}=1.0$ at $S_\xi=\sigma^2_8=0.64$ (the
triangle). Clearly, the larger the value of 
$\delta_{\rm env}$, the more likely
the Brownian motion will hit the barrier at smaller $S$.
 This is what we see in
Fig.~\ref{fig:fS}, which shows the conditional distribution
$f(\delta_c(S,\delta_{\rm env}),S~|~\delta_{\rm env},S_\xi)$ for
different values of $\delta_{\rm env}$.

For comparison we also show the corresponding results for the
$\Lambda$CDM model using the dashed curves in Fig.~\ref{fig:fS}. Note
that the solid curves are always higher than the dashed ones for
smaller $S$ and lower for bigger $S$. This is because in the chameleon
model the barrier is generally lower and the Brownian motion is 
likely to cross it for the first time  at smaller $S$.

\subsubsection{Integrating over the Environment Distribution}

To get the final first crossing distribution of the moving barrier, we must integrate over all environments. The distribution of $\delta_{\rm env}$, denoted as $q(\delta_{\rm env},\delta_{\rm sc},S_\xi)$, in which $\delta_{\rm sc}$ is the critical overdensity for the spherical collapse in the $\Lambda$CDM model\footnote{Remember again that the evolution of the environment is assumed to be governed by the $\Lambda$CDM model.}, is simply the probability that the Brownian motion passes $\delta_{\rm env}$ at $S_\xi$ and never exceeds $\delta_{\rm sc}$ for $S<S_\xi$ (because otherwise the environment itself has collapsed already). This has been derived by \cite{bcek}:
\begin{eqnarray}\label{eq:q}
q(\delta_{\rm env},\delta_{\rm sc},S_\xi) &=& \frac{1}{\sqrt{2\pi S_\xi}}\exp\left[-\frac{\delta^2_{\rm env}}{2S_\xi}\right]\nonumber\\
&&-\frac{1}{\sqrt{2\pi S_\xi}}\exp\left[-\frac{\left(\delta_{\rm env}-2\delta_{\rm sc}\right)^2}{2S_\xi}\right],
\end{eqnarray}
for $\delta_{\rm env}\leq\delta_{\rm sc}$ and $0$ otherwise.

Then the environment-averaged first crossing distribution will be
\begin{eqnarray}\label{eq:f_ave}
f_{\rm ave}(S) &=& \int^{\delta_{\rm sc}}_{-\infty}q\times f(\delta_c(S,\delta_{\rm env}),S~|~\delta_{\rm env},S_\xi)d\delta_{\rm env}.
\end{eqnarray}
In the special case where the barrier is flat, $\delta_{c}(S,\delta_{\rm env})=\delta_{\rm sc}$, $f(\delta_c(S,\delta_{\rm env}),S~|~\delta_{\rm env},S_\xi)$ is known analytically as
\begin{eqnarray}
f &=& \frac{\delta_{\rm sc}-\delta_{\rm env}}{\sqrt{2\pi}\left(S-S_\xi\right)^{3/2}}\exp\left[-\frac{\left(\delta_{\rm sc}-\delta_{\rm env}\right)^2}{2\left(S-S_\xi\right)}\right],
\end{eqnarray}
and the integration in Eq.~(\ref{eq:f_ave}) can be performed exactly to obtain
\begin{eqnarray}
f_{ave}(S) &=& \frac{1}{\sqrt{2\pi S}}\frac{\delta_{\rm sc}}{S}\exp\left[-\frac{\delta^2_{\rm sc}}{2S}\right],
\end{eqnarray}
which is just the unconditional first crossing distribution for a
constant barrier $\delta_{\rm sc}$ at $S$. This is as expected,
because the collapse does not depend on the environment.

\begin{figure}
\includegraphics[width=88mm]{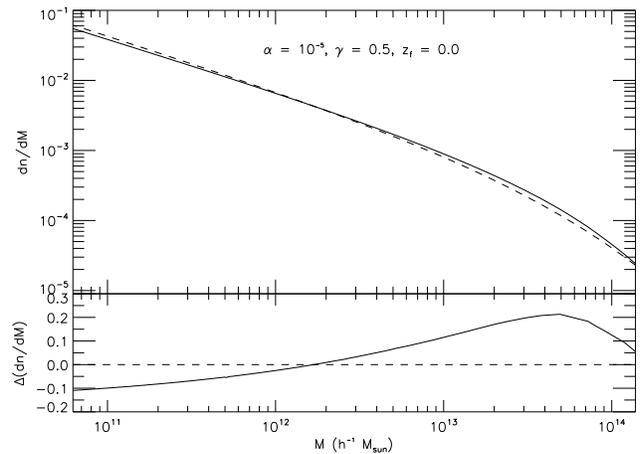}
\caption{{\it Upper panel}: the mass functions for the chameleon (solid curve) and $\Lambda$CDM (dashed curve) models. Some physical parameters are shown in the figure, and others include $\Omega_m=0.24$. {\it Lower panel}: the fractional differences between the two mass functions (solid curve); the dashed line is identically zero and is shown as a reference.}
\label{fig:mf}
\end{figure}

In general cases with environment-dependent collapse, $f_{\rm ave}(S)$
must be computed numerically. Indeed, in Eq.~(\ref{eq:f_ave}) both
$q(\delta_{\rm env},\delta_{\rm sc},S_\xi)$ and
$f(\delta_c(S,\delta_{\rm env}),S~|~\delta_{\rm env},S_\xi)$ differ
from the flat-barrier case. The distribution $f$ has been discussed
above (cf.~Fig.~\ref{fig:fS}). The distribution $q$ should, in
principle, be calculated for the chameleon model numerically, but we
choose to use the $\Lambda$CDM result Eq.~(\ref{eq:q}) for the
following reasons: recall that $q(\delta_{\rm env},\delta_{\rm
  sc},S_\xi)$ is the probability that the Brownian motion starts off
at the origin, never hits the constant barrier $\delta_{\rm sc}$
before $S_\xi$ and goes through $\delta_{\rm env}$ at $S_\xi$. To
estimate its difference from the true value in the chameleon model, we
replace the $\delta_{\rm sc}$ in Eq.~(\ref{eq:q}) with
$\delta_c(S_\xi)$ for the $\delta_{\rm env}$ values in the figures,
and find that the change of $q$ is at percent and subpercent
level\footnote{Noe that this is an {\it upper} bound of the error of
  using Eq.~(\ref{eq:q}), because the barrier does not stay at
  $\delta_c(S_\xi)$ for all $S\in[0,S_\xi]$ but rather decreases from
  $\delta_{\rm sc}$ at $S=0$ to it at $S=S_\xi$,}, which is not
surprising given that $\delta_{c}(S\leq S_\xi)$ is very close to
$\delta_{\rm sc}$ (cf.~Fig.~\ref{fig:bm}) (the approximation will be
even better for higher redshift cf.~Fig.~\ref{fig:dc_envb}). A better
approximation would be to assume
$\delta_{c}(S)\approx\delta_{\sc}-\beta S$ where $\beta$ is some
constant, but here we do not see the necessity for doing this.

Using Eq.~(\ref{eq:q}), we perform the integral in
Eq.~(\ref{eq:f_ave}) using  Gaussian quadrature. We checked
the accuracy of this method by applying it to the flat-barrier case
and find that the agreement with the exact solution is excellent. The
halo mass function is related to the averaged first-crossing
distribution $f_{\rm ave}(S)$ by
\begin{eqnarray}
\frac{dn}{dM}dM = \frac{\bar{\rho}_m}{M}f_{\rm ave}(S)\left|\frac{dS}{dM}\right|dM,\nonumber
\end{eqnarray}
and we have plotted in Fig.~\ref{fig:mf} the function $dn(M,z=0)/dM$ for
both the chameleon (solid curves) and the $\Lambda$CDM (dashed curve)
models. As can be seen clearly, the fifth force results in  more
massive halos than in the $\Lambda$CDM model,  but in compensation
there are fewer low mass halos ($M <
10^{12}h^{-1}M_{\bigodot}$) in the chameleon model.

To see the difference more clearly, we have also plotted the
fractional difference between the chameleon and $\Lambda$CDM mass
functions in the lower panel of Fig.~\ref{fig:mf}. This shows that the
increase of $n(M)$ is largest for halos in the mass range $10^{13} <
M/\left(h^{-1}M_{\bigodot}\right) < 10^{14}$. For high mass halos the
fifth force is strongly suppressed and its effect on the mass 
function is smaller, as
expected. Because a larger fraction of the total mass has been
assembled in high mass halos, fewer small isolated halos  survive the
merger and accretion process\footnote{We want to emphasise that the
  result here is not directly comparable with that obtained in
  \cite{lz2010} using full $N$-body simulations, because there
  subhalos are also counted.}.

\begin{figure*}
\includegraphics[width=139mm]{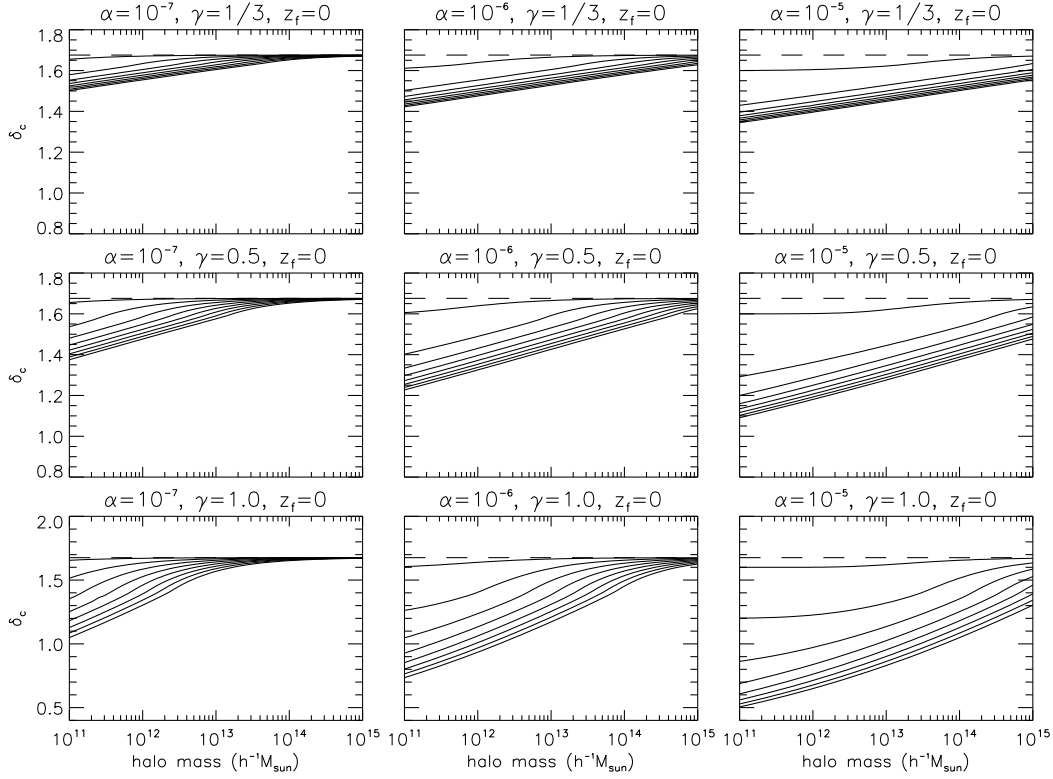}
\caption{The critical $\Lambda$CDM-linearly-evolved overdensity for spherical collapse at $z_f=0$ as a function of the mass enclosed and the environment $\delta_{\rm env}$. The physical parameters $\alpha, \gamma$ are indicated beside each panel. In each panel the solid curves from top to bottom are respectively for $\delta_{\rm env}=1.6, 1.2, 0.8, 0.4, 0.0, -0.4, -0.8$ and $-1.2$. The dashed line is the result for $\Lambda$CDM model.}
\label{fig:dc_env_z0}
\end{figure*}

\begin{figure*}
\includegraphics[width=139mm]{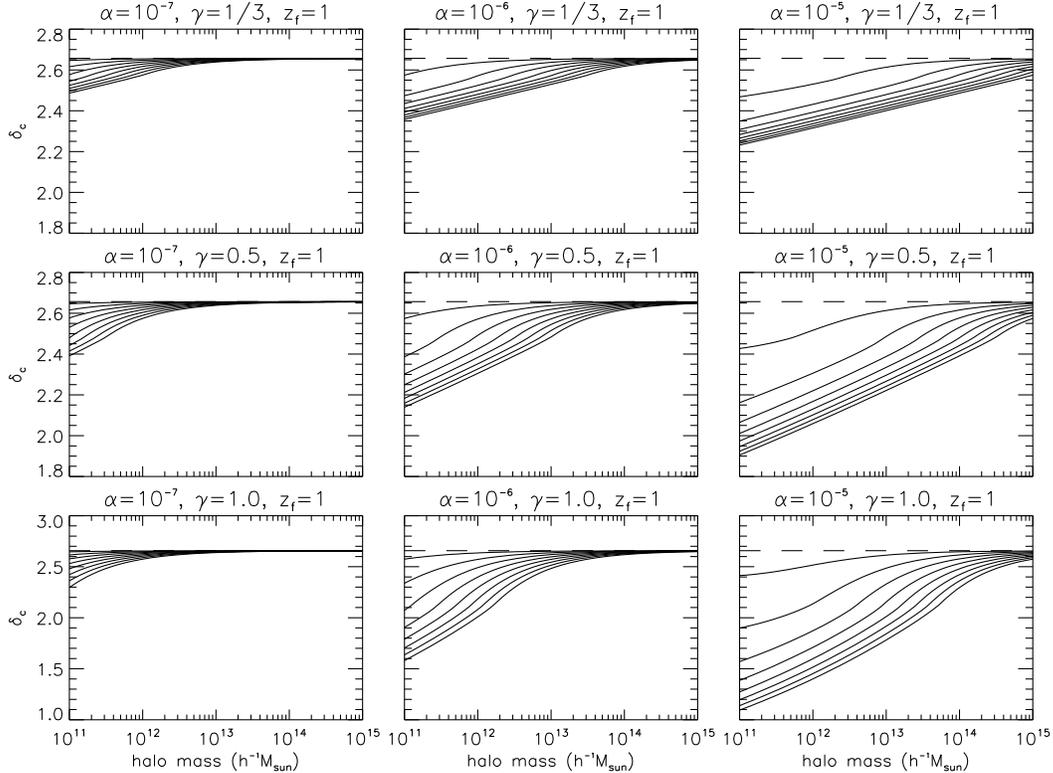}
\caption{The same as Fig.~\ref{fig:dc_env_z1}, but for collapsing redshift $z_f=1$. In each panel the solid curves from top to bottom are respectively for $\delta_{\rm env}=2.4, 1.8, 1.2, 0.6, 0.0, -0.6, -1.2$ and $-1.8$.}
\label{fig:dc_env_z1}
\end{figure*}

\begin{figure*}
\includegraphics[width=150mm]{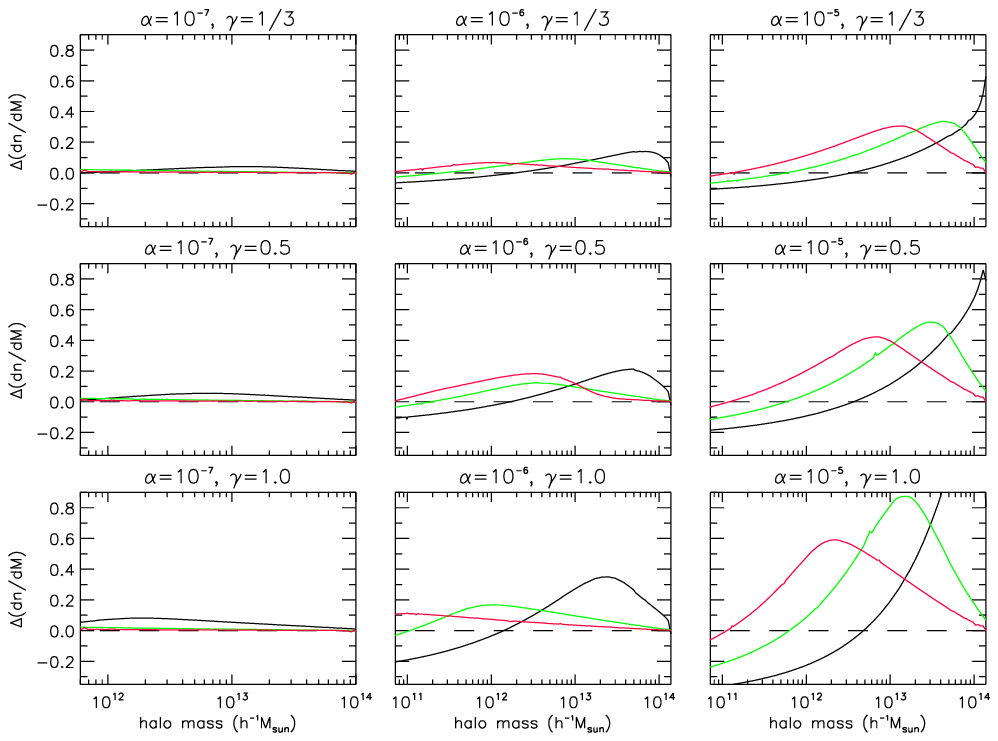}
\caption{(Colour online) The fractional difference of $dn/dM$ between the chameleon (the scalar field parameters $\alpha$ and $\gamma$ are indicated beside each panel) and $\Lambda$CDM models, at three redshifts $0$ (black curves), $1$ (green curves) and $2$ (red curves). The result for $\Lambda$CDM is plotted as the dashed line for reference.}
\label{fig:dmf_z}
\end{figure*}

Note that the effects of the fifth force are suppressed for high mass
halos,  not only because the halos are  efficient at screening that
force themselves, but also because they are more likely to reside in
dense environments. More explicitly, the probability distribution of
$\delta_{\rm env}$ at $S=S_\xi$, given that the Brownian motion goes
through $\delta\sim\delta_c(S)$ at $S$ (where it is about to cross the
barrier), is
\begin{eqnarray}
p(\delta_{\rm env}~|~S,\delta) = \frac{1}{\sqrt{2\pi\frac{S_\xi}{S}(S-S_\xi)}}\exp\left[-\frac{\left(\delta_{\rm env}-\frac{S_\xi}{S}\delta\right)^2}{2\frac{S_\xi}{S}(S-S_\xi)}\right].\nonumber
\end{eqnarray}
For high mass halos, $S$ is close to $S_\xi$ and this distribution strongly peaks at $\delta_{\rm env}\sim S_{\xi}\delta/S\sim\delta$. The results, of course, are consistent with our intuitive understanding about the chameleon effect.

\section{Applications}

\label{sect:applications}

The key new concept in our extended excursion set model is the
specification of the environment in terms of two parameters
$(S_\xi,\delta_{\rm env})$: the environment determines how spherical
collapse in the chameleon model is modified compared to $\Lambda$CDM,
and also means that we have to use conditional distribution of the
first crossing rather than the unconditional distriubtion, as in the
conventional excursion set approach, to compute observables such as
the mass function of non-linear structures.

The use of the conditional first-crossing distribution is not
new. \cite{mw1996}, for example, used it to study the bias between the
halo number density and the dark matter density fields in the
$\Lambda$CDM model. In the chameleon model, both bias and the mass
function must  be computed using the conditional first-crossing
distribution. In fact, computation of the mass
function is more complicated since we need
to average over the probability distribution of environments.

The methods introduced here can be used to study, for example, the
formation redshift of halos $z_f$ and their dependence on the
parameters $\alpha$ and $\gamma$ describing the chameleon mechanism,
$z_f$.  The most difficult step in such a calculation
is the computation of the moving and environment dependent barrier
  $\delta_c(S,\delta_{\rm env})$. Nevertheless, the computations
are very much faster that $N$-body simulations and so we can
explore large regions of parameter space rapidly.


\begin{table}
\label{table:param} \caption{The parameters $\alpha$ and $\beta$ for the 9 chameleon models studied in this section. The $\Lambda$CDM paradigm corresponds to $\alpha=\beta=0$. We assume $\Omega_m=0.24$ and $\Omega_{\Lambda}=0.76$.}
\begin{tabular}{@{}lcccccccccc}
\hline\hline
$\lg(\alpha)$ & $-7$ & $-7$ & $-7$ & $-6$ & $-6$ & $-6$ & $-5$ & $-5$ & $-5$\\
$\beta$ & $1/3$ & $1/2$ & $1$ & $1/3$ & $1/2$ & $1$ & $1/3$ & $1/2$ & $1/1$\\
\hline
\end{tabular}
\end{table}

Let us consider the mass functions of the chameleon models with model
parameters as specified in Table~\ref{table:param}.
Fig.~\ref{fig:dc_env_z0} shows the critical density for a spherical
overdensity to collapse at $z_f=0$ as a function of the enclosed mass
$M$ and environment $\delta_{\rm env}$. As expected, the collapse
threshold is lower in all chameleon models because the fifth force,
however weak, is always attractive and boosts the collapse. For
smaller $\alpha$ (left column, $\alpha=10^{-7}$), the difference from
the $\Lambda$CDM prediction is small, especially for the largest
overdensities because the fifth force is more strongly suppressed in these
systems, as discussed in the previous section.  On the other hand, for
large $\alpha$ (middle and right columns, $\alpha=10^{-6}$ and
$10^{-5}$), the deviation from $\Lambda$CDM is much larger, even for
the largest overdensities. Increasing $\gamma$ will strengthen the
fifth force and therefore also lower the collapse threshold.

Fig.~\ref{fig:dc_env_z1} is equivalent to 
Fig.~\ref{fig:dc_env_z0}, but for spherical overdensities which
collapse at $z_f=1$. Because the matter density is higher at
higher redshift, the fifth force is more strongly suppressed and 
hence the deviation from $\Lambda$CDM is smaller.

Finally, we have plotted the effect of a chameleon-type fifth force on
the dark matter halo mass functions in Fig.~\ref{fig:dmf_z}. For
clarity Fig.~\ref{fig:dmf_z} shows the fractional change of the
quantity $dn/dM$, where $n(M)$ is the halo mass function, with respect
to the $\Lambda$CDM prediction, at three redshifts $z=0,1,2$
respectively. For $\alpha=10^{-7}$, the fifth force is strongly
suppressed and the fractional change of $dn/dM$ is less than $10$\%,
even for $\gamma=1$ and $z=0$.

For $\alpha=10^{-6}$, the fifth force is less suppressed and the
fractional change of $dn/dM$ at $z=0$ could be up to $\sim15$\% (for
$\gamma=1/3$) or even $\sim40$\% (for $\gamma=1$), showing interesting
and potentially observable effects. The deviation at early times
is mainly restricted to lower mass halos.  At later
times, massive halos also start to feel the fifth force, and a 
deviation is seen  at higher halo masses.
 With $\alpha=10^{-5}$, the qualitative features mentioned
above all remain, but the deviation from $\Lambda$CDM is much
stronger, up to $\sim60$\% for $\gamma=1/3$ and more than $100$\% for
$\gamma=1$ at $z=0$. 

Figs.~\ref{fig:dc_env_z0}-\ref{fig:dmf_z} show that some choices of parameters can lead
to large deviations in the abundances of non-linear objects
compared to the $\Lambda$CDM model. The tightest constraints on the
parameters $\alpha$ and $\beta$ would probably come from 
number counts and number densities of well characterised galaxy
cluster samples intermediate redshifts, $z < 1$. Such samples
are becoming available from combined Sunayev-Zeldovich/X-ray measurements
from Planck \citep{planck} and SPT \citep{spt}. A detailed comparison of observations
with our model is beyond the scope of this paper. 

One question one might ask is if the chameleon model could be used to
produce more halos at very early times, say $z>6$, which might
ease problems in reionizing the intergalactic medium at early
times.  It has been argued (\cite{hkk2010})
 that a fifth force might boost hierarchical structure
formation leading to enhanced production of UV photons
at  early times. Unfortunately, the chameleon-type fifth force 
is   strongly suppressed at earlier times.  Fig.~\ref{fig:dmf_z} shows
this up to $z=2$, and for $z>6$ the deviation from $\Lambda$CDM is
even smaller.

\section{Summary and Conclusions}

\label{sect:summary}

To summarise, in this paper we have presented 
an extension of the standard excursion set theory 
so that it can be used to study structure-formation scenarios 
which are environmentally dependent. 
Our method separates the calculation into  two steps:
\begin{enumerate}
\item compute the collapse of the spherical overdensity in a given environment;
\item compute the probability that the 
spherical overdensity is located in the specified environment, 
and average over the distribution of environments.
\end{enumerate} 
For (i) we have proposed a simplified model, which is a
generalisation of the usual spherical collapse model to the case
in which the overdensity evolves inside an evolving environment.
For (ii) we have derived an approximation to the environment distribution, 
and shown  how to compute the averaged first-crossing
distribution, which is closely related to the halo mass functions.

As a working example, we have applied  the method to 
the chameleon model. Our numerical results agree with 
how we expect the chameleon effect to behave as a function
of the model parameters. We have concentrated here on
the collapse redshift and mass functions of virialized
objects. These predictions could be used in conjunction
with forthcoming data to set constraints on the model parameters.

As in the standard excursion set theory for the cold dark matter
model, it is straightforward to generalise the excursion
set approach to compute other observables, for example
the formation of voids \citep{sv2004}  and the merger history of 
halos \citep{lc1993}. In conjunction with the halo model, our method can be
used to predict the non-linear matter power spectrum as well. 

Structure formation scenarios with strong environment dependence have
become more and more popular recently.  Newtonian gravity
has been tested to high precision in our
local environment, but deviations may be significant on cosmological
scales,  perhaps as a result of a chameleon-like mechanism. 
The methods presented here offer a  faster alternative
to  $N$-body simulations, enabling a much wider range
of models to be confronted with observations. \tcr{Meanwhile, the analytic formulae presented here enable a clear track of the underlying physics, such as in which ways the chameleon effect modifies the structure formation.} We hope
that these methods will contribute towards a better
understanding of the nature of dark energy.

\section*{Acknowledgments}

BL is supported by Queens' College, the Department of Applied Mathematics and Theoretical Physics of University of Cambridge, \tcr{and the Royal Astronomical Society.}

%

\label{lastpage}

\end{document}